\documentstyle[12pt,epsfig]{article}

\newlength{\dinwidth}
\newlength{\dinmargin}
\def\lapproxeq{\lower .7ex\hbox{$\;\stackrel{\textstyle
<}{\sim}\;$}}
\def\gapproxeq{\lower .7ex\hbox{$\;\stackrel{\textstyle
>}{\sim}\;$}}
\def\gtrsim{\lower .7ex\hbox{$\;\stackrel{\textstyle
>}{\sim}\;$}}
\def\lesim{\lower .7ex\hbox{$\;\stackrel{\textstyle
<}{\sim}\;$}}

\def\be{\begin{equation}}
\def\ee{\end{equation}}
\def\bea{\begin{eqnarray}}
\def\eea{\end{eqnarray}}

\def\bb{b\bar{b}}

\def\GeV{\rm GeV}

\newcommand{\eq}[1]{(\ref{eq:#1})}
\newcommand{\bi}[1]{\bibitem{a#1}}

\begin{document}
%\titlepage
\begin{flushright}
IPPP/03/35 \\
DCPT/03/70 \\
29 September 2003 \\

\end{flushright}

\vspace*{0.5cm}

\begin{center}
{\Large \bf Central exclusive diffractive production as a\\[1ex] spin--parity analyser: from hadrons to Higgs}

\vspace*{1cm}
\textsc{A.B.~Kaidalov$^{a,b}$, V.A.~Khoze$^{a,c}$, A.D. Martin$^a$ and M.G. Ryskin$^{a,c}$} \\

\vspace*{0.5cm} $^a$ Department of Physics and Institute for
Particle Physics Phenomenology, \\
University of Durham, DH1 3LE, UK \\
$^b$ Institute of Theoretical and Experimental Physics, Moscow, 117259, Russia\\
$^c$ Petersburg Nuclear Physics Institute, Gatchina,
St.~Petersburg, 188300, Russia \\

\end{center}

\vspace*{0.5cm}

\begin{abstract}
We present the general rules for double-Reggeon production of objects with different spins and parities. The
existing experimental information on resonance production in these central exclusive diffractive processes is
discussed. The absorptive corrections are calculated and found to depend strongly on the quantum numbers of the
produced states. The central exclusive diffractive production of $0^+$ and $0^-$ Higgs bosons is studied as an
illustrative topical example of the use of the general rules. The signal for diffractive $0^+$ and $0^-$ Higgs
production at the LHC is evaluated using, as an example, the minimal supersymmetric model, with large $\tan\beta$.
\end{abstract}

\section{Introduction}

\begin{figure}
\begin{center}
\includegraphics[height=5cm]{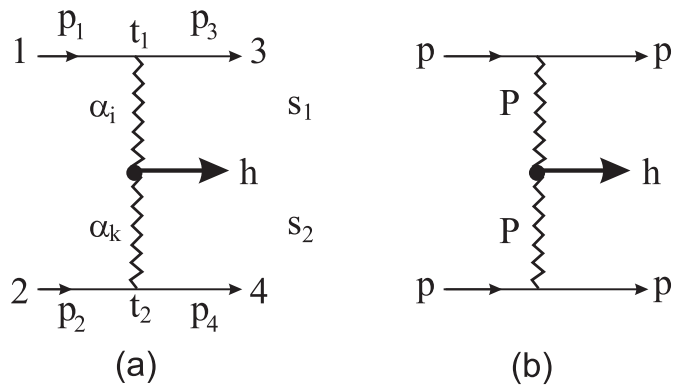}
\caption{(a)~The central production of a state $h$ by double-Reggeon exchange.~~(b)~The double-Pomeron exchange
contribution to $pp\to p + h + p$, which dominates at high energies, where the $+$~signs are used to indicate the
presence of Pomeron-induced rapidity gaps.\label{fig:1}}
\end{center}
\end{figure}

It is always a challenge to measure the quantum numbers of new states, particularly their spin and parity. We may
measure the specific characteristics of given decay channels or angular correlations of the accompanying particles
in the production process, especially the correlations between the outgoing protons in the central exclusive
production process, $pp\to p + h + p$, shown in Fig.~\ref{fig:1}(a).
The advantage of the latter approach is that it offers the possibility, not only to separate different states by
accurately measuring the missing mass, but also to distinguish between scalar and pseudoscalar new heavy objects,
which is difficult from studying their decay products. In this paper we begin by studying the general implications
of applying Reggeon techniques to describe such exclusive processes. At very high energies, and in the central
region ($x_F\simeq0$), the double-Pomeron process, Fig~1(b), should give the dominant contribution.

The theory of double-Reggeon (and multi-Reggeon) exchanges was developed long ago~\cite{a1}. However the revival
of interest in these processes is related to the new effects observed in the central production of resonances by
the WA102 experiment~\cite{a2} in the reaction $pp\to pX^0p$, and to the proposal to look for the Higgs boson and
other new particles in double-Pomeron-exchange processes, see, for example,~\cite{BL}--\cite{INC}. Indeed it will
be one of the main challenges of the LHC to identify the nature (including the spin--parity) of newly-discovered
heavy objects. It appears to be very hard to find a spin--parity analyser using conventional approaches.

Models for double-Pomeron-exchange production of hadrons with different quantum numbers have been developed in
recent years~\cite{a4}--\cite{a8}. However in some papers~\cite{a5,a5a} the formulas of Reggeon theory were not
fully consistent, while some results of the others follow simply from general rules of the Reggeon approach. In
Section~\ref{sec:2} we first consider these rules and compare them with experimental observations~\cite{a2}, and
with the results of the phenomenological analysis performed in Ref.~\cite{a7}. Also the dynamics for the
Pomeron--Pomeron--particle vertices is discussed.

\begin{figure}
\begin{center}
\includegraphics[height=5cm]{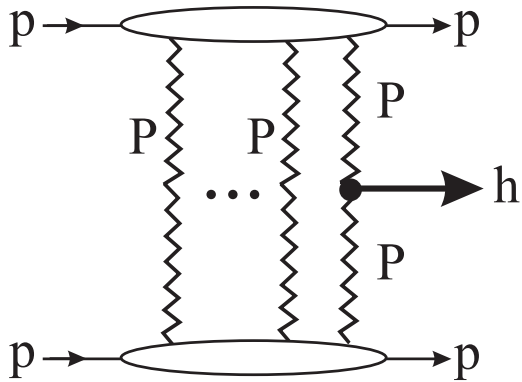}
\caption{Unitarity or rescattering corrections to the exclusive diffractive process $pp\to p + h + p$.
\label{fig:2}}
\end{center}
\end{figure}

In Section~\ref{sec:3} we illustrate how the general behaviour is distorted by the dynamics of the process, using
$h(0^\pm)$ exclusive diffractive production as an example. Apart from subsection~3.1, where we discuss the
uncertainties in the predicted cross sections, this section neglects the absorptive or unitary corrections.
However at high energies these corrections are important (see, for example, Refs.~\cite{Bj}--\cite{a17}). They
correspond to the diagrams of Fig.~\ref{fig:2}, and can be calculated using the Reggeon diagram
technique~\cite{a10}.
It will be shown in Section~\ref{sec:4} that the inclusion of these diagrams leads not only to a decrease of the
cross sections of the double-Pomeron processes, but also to significant modifications of the angular correlations
between the outgoing (forward) protons. Moreover, the magnitude of these absorptive corrections depends on the
quantum numbers of the produced state $h$.

In Section~\ref{sec:4} we illustrate the results using the important topical example of the double-Pomeron
production of heavy bosons. In particular we compare the production of a Standard Model Higgs boson with
spin-parity $J^P=0^+$, with that for a $0^-$ Higgs\footnote{For convenience of presentation we will denote this
particle $h(0^-)$, rather than the conventional notation $A(0^-)$.} which appears in various extensions of the
Standard Model, in particular in a supersymmetric extension. In Section~\ref{sec:5}, we consider the consequences
of this approach to investigations of the Higgs sector at the LHC. For illustration we evaluate the exclusive
cross sections using the minimal supersymmetric model\footnote{For a recent review see, for example,
Ref.~\cite{CH}.} (MSSM) with large $\tan\beta$; a domain in which, for $m(0^-)\lesim 200~\GeV$, the conventional
searches at the LHC will face difficulties to discriminate between the different Higgs states and to determine
their masses. This is especially true in the so-called ``intense coupling limit'', $m_h\sim m_A\sim m_H\sim {\cal
O}(100~\GeV)$~\cite{EB}. As a specific example we calculate the event rates for the exclusive central production
of mass $115~\GeV$ $0^\pm$ bosons at the LHC.

\section{Exclusive diffractive production: general rules} \label{sec:2}

Here we study the general rules for the amplitudes of the exclusive diffractive process
\be 1+2\ \to\ 3+h+4, \label{eq:adder} \ee
shown in Fig.~\ref{fig:1}, where $1,\dots,4$ are hadrons, and where the centrally produced particle $h$ has spin
and parity $J^P$. We show that the production process has characteristic features, that depend on the value of
$J^P$, which follow from general principles.

To begin, we assume that all the particles are spinless. Then, at high energies and small momentum transfer,
\be\begin{array}{l @{\qquad} l} s_1 = (p_3+p_h)^2 \gg s_0,  &  s_2 = (p_4 + p_h)^2 \gg s_0  \\
t_1\simeq -p_{3\perp}^2 \lapproxeq s_0, &  t_2 \simeq -p_{4\perp}^2 \lapproxeq s_0,  \label{eq:bear}
\end{array}\ee
the amplitude can be written in the form~\cite{a1}
\be T_{12}^{3h4}(s_1,s_2,\vec p_{3\perp}, \vec p_{4\perp}) = \sum_{i,k}
g_{13}(t_1)g_{24}(t_2)\left(\frac{s_1}{s_0}\right)^{\alpha_i(t_1)}\left(\frac{s_2}{s_0}\right)^{\alpha_k(t_2)}
\eta(\alpha_i(t_1))\eta(\alpha_k(t_2)) g_{ik}^h(t_1,t_2,\phi), \label{eq:cat} \ee
where $s_0 = 1~\GeV^2$, $\phi$ is the angle between the transverse momenta $\vec p_{3\perp}$ and $\vec p_{4\perp}$
of the outgoing protons and
\be \eta(\alpha_i(t)) = -\left(\frac{1+\sigma_i e^{-i\pi\alpha_i(t)}}{\sin\pi\alpha_i(t)}\right) \label{eq:dog}
\ee
is the signature factor for Regge pole $i$ with trajectory $\alpha_i(t)$ and signature $\sigma_i = \pm1$. The
vertex factors $g_{13}(t_1)$ and $g_{24}(t_2)$ describe the $13-\alpha_i$ and $24-\alpha_k$ couplings
respectively, while $g_{ik}^h$ describes the transition $\alpha_i\alpha_k\to h$. Note that $g_{ik}^h$ depends, in
general, on all the scalars that can be formed from the vectors which enter the vertex. Moreover, in the case of
Reggeons, the longitudinal and transverse components of their momenta act as two different vectors~\cite{a10}. In
our case, where the mass of the boson $h$ is fixed, it is enough to keep the transverse momenta $\vec p_{3\perp}$
and $\vec p_{4\perp}$, and the unit vector $\vec n_0$ in the direction of the colliding hadrons. Unlike $g_{13}$
and $g_{24}$, the function $g_{ik}^h$ may be complex. In the Regge domain~\eq{bear},
\be s_1s_2 = s(m_h^2 + p_{h\perp}^2), \label{eq:elephant} \ee
where $s = (p_1 + p_2)^2$.

When spin is included, the process is described by helicity amplitudes, each of which has a double-Regge behaviour
as in \eq{cat}~\cite{a11}.
\be T_{\lambda_1\lambda_2}^{\lambda_3\lambda_h\lambda_4}(s_1,s_2,t_1,t_2,\phi) =
\sum_{i,k}g_{\lambda_1\lambda_3}(t_1)g_{\lambda_2\lambda_4}(t_2)
\left(\frac{s_1}{s_0}\right)^{\alpha_i(t_1)}\left(\frac{s_2}{s_0}\right)^{\alpha_k(t_2)}
\eta(\alpha_i(t_1))\eta(\alpha_k(t_2))g_{ik}^{\lambda_h}(t_1,t_2,\phi)  \label{eq:fox} \ee
The relations between the vertex couplings for different helicities, due to conservation of parity and other
quantum numbers, are the same as for $2\to2$ reaction~\cite{a12}. For example,
\be g_{\lambda_1\lambda_3}(t) = (-1)^{\lambda_1-\lambda_3}\xi_1\: g_{-\lambda_1-\lambda_3}(t), \label{eq:giraffe}
\ee
with
\be \xi_1 = \eta_1\eta_3(-1)^{S_1-S_3}P_i\sigma_i, \label{eq:hedgehog} \ee
where $S_1$ ($S_3$) and $\eta_1$ ($\eta_3$) are the spin and parity of the particle 1 (3) respectively, and
$P_i,\sigma_i$ are the parity and signature of the Reggeon $i$. The vertices behave as \cite{a12,a13},
\be g_{\lambda_1\lambda_3}(t) \sim (-t)^{\left| \lambda_1 - \lambda_3\right|/2},\quad {\rm as}\ t\to 0.
\label{eq:iguana} \ee
Note that relation \eq{giraffe} depends only on the product $P_i\sigma_i$ and thus the model-independent spin
structure of the vertices $g_{\lambda_1\lambda_3}(g_{\lambda_2\lambda_4})$ is the same for all Reggeons with the
same product $P_i\sigma_i$.

Below we will be interested mainly in the spin structure of the central vertex $g_{ik}^{\lambda_h}(t_1,t_2,\phi)$.
It can be written in the form~\cite{a11}
\be g_{ik}^{\lambda_h} = \sum_{m_2 = -\infty}^\infty e^{im_2\phi}\: \gamma_{m_1m_2}^{\lambda_h},\quad {\rm with}\
m_1 + m_2 = \lambda_h, \label{eq:jackal} \ee
where $m_2$ has the meaning of the projection of the angular momentum $j_k$ of Reggeon $k$ (analytically continued
from all angular momenta $j_k$ in the $t_1,t_2$-channels). Now invariance under parity leads to the
relation~\cite{a11}
\be \gamma_{m_1m_2}^{\lambda_h} = (-1)^{\lambda_h}\xi_3\: \gamma_{-m_1-m_2}^{-\lambda_h}, \label{eq:koala} \ee
where
\be \xi_3 = \eta_h(-1)^{S_h}P_i\sigma_iP_k\sigma_k. \label{eq:lion} \ee
Thus the spin structure of the central vertex depends only on the product of the naturalities (that is the
parities and signatures) of particle $h$ and the exchanged Reggeons\footnote{It can be shown that similar formulae
are also valid for photon--photon fusion processes.}. The behaviour of $\gamma_{m_1m_2}^{\lambda_h}$ for small
$t_1,t_2$ is given by the formula~\cite{a11}
\be \gamma_{m_1m_2}^{\lambda_h} \sim (-t_1)^{|m_1|/2} (-t_2)^{|m_2|/2},\quad{\rm with}\ m_1+m_2 = \lambda_h.
\label{eq:mouse} \ee
Note that all values of $m_2$ ($m_1$) enter~\eq{jackal}, but, due to~\eq{mouse}, for small $t_1,t_2$ it is enough
to consider the lowest values of $m_2$ ($m_1$) consistent with \eq{koala}.

It is often convenient to write the spin structure of the amplitudes in terms of the characteristic 3-vectors of
the problem. Such a representation is closely related to the helicity amplitudes discussed above~\cite{a13}, but
the formulas become more transparent. In this case the central vertex $g_{ik}^h$ is written as a scalar (or
pseudoscalar) function (depending on the product $\eta_h(-1)^{S_h}\sigma_iP_i\sigma_kP_k$) of the vectors $\vec
p_{3\perp},\vec p_{4\perp}$ and the spin vectors (tensors) of particle $h$.

Let us consider particular examples for the spin-parity $J^P$ of $h$, in each case taking\linebreak
$\sigma_iP_i\sigma_kP_k = +1 $ as for double-Pomeron exchange.

\begin{itemize}
\item[(a)] $J^P(h) = 0^+$

For a scalar particle $h$, the vertex coupling is simply
\be g_{ik}^h = f_{0^+}(p_{3\perp}^2, p_{4\perp}^2, \vec p_{3\perp}\!\cdot\!\vec p_{4\perp}),
\label{eq:nightingale} \ee
where $f_{0^+}$ is a function of the scalar variables which can be formed from the transverse momenta $\vec
p_{3\perp}$ and $£\vec p_{4\perp}$ of the outgoing protons. When $\vec p_{3\perp}$ or $\vec p_{4\perp}\to 0$, this
function in general tends to some constant $f(0,0,0)$. In order to obtain further information on the structure of
this function, extra dynamical input is needed (see Section~\ref{sec:3}).

\item[(b)] $J^P(h) = 0^-$

For the central production of a pseudoscalar particle, the vertex factor takes the form
\be g_{ik}^h = f_{0^-}(p_{3\perp}^2, p_{4\perp}^2, \vec p_{3\perp}\!\cdot\!\vec p_{4\perp})\: \varepsilon_{ikl}
(p_{3\perp})_i(p_{4\perp})_k (n_0)_l, \label{eq:ocelot} \ee
where $\vec n_0$ is the unit vector in the direction of the colliding hadrons (in the c.m.s.). In this case,
according to~\eq{koala}, all amplitudes with $m_1,m_2=0$ are zero, and so $|m_1| = |m_2| = 1$ are the leading
terms. According to~\eq{mouse} this corresponds to helicity amplitudes which are proportional to
$(-t_1)^{\frac{1}{2}}(-t_2)^{\frac{1}{2}}$ for small $t$. Thus the cross section behaves as $\sim|t_1||t_2|$. Also
the angular distribution contains a factor $\sin^2\phi$, which is evident from either~\eq{ocelot} or \eq{jackal}
and \eq{koala}. Again the function $f_{0^-}$ is not predicted from general principles.

Note that the characteristic $\sin^2\phi$ dependence of the angular distribution, and the $t$-behaviour at small
$t$, which are observed by the WA102 Collaboration for $\eta$, and $\eta'$ production~\cite{a2}, are direct
consequences of the general properties of the double-Regge-exchange amplitudes. This behaviour is valid not only
for double-Pomeron exchange, but also for $P\!f, f\!f, \rho\rho, A_2A_2, \omega\omega,\dots$ exchanges. Under the
interchange of the Reggeons, $i\leftrightarrow k$,
\be (n_0)_l \to -(n_0)_l \label{eq:panda} \ee
and hence if the Reggeons are the same, $i=k$, the function $f_{0^-}$ should be symmetric under the interchange
$3\leftrightarrow 4$.

\item[(c)] $J^P(h) = 1^+$

For the production of an axial vector state, both $\lambda_h = 0$ and $\lambda_h = \pm1$ components are present,
and the vertex factor can be written as
\be g_{ik}^h = f_{1^+}^0\: \varepsilon_{ikl} (p_{3\perp})_i(p_{4\perp})_k e_l + \left( f_{1^+}^1\:(p_{3\perp})_i +
\tilde f_{1^+}^1\:(p_{4\perp})_i\right) \varepsilon_{ikl}(n_0)_ke_l, \label{eq:quail} \ee
where the $f^i$ are functions of the scalar variables $p_{3\perp}^2, p_{4\perp}^2$ and $\vec p_{3\perp} \cdot \vec
p_{4\perp}$, and where $\vec e$ is the polarization vector of the $1^+$ meson. If both of the exchanged Reggeons
are the same ($i=k$), then the function $g_{ik}^h$ is symmetric under the interchange $3\leftrightarrow 4$. As a
consequence, for small $\vec  p_{i\perp}$,
\be f_{1^+}^0 \sim (p_{3\perp}^2 - p_{4\perp}^2), \qquad f_{1^+}^1 = -\tilde f_{1^+}^1. \label{eq:rat} \ee
\end{itemize}

The form of the vertex factors $g_{ik}^h$ for the central production of states $h$ of higher spin can readily be
constructed using equations~\eq{jackal}--\eq{lion}.

It is interesting to note that the structure of the vertex factors given in \eq{nightingale}--\eq{rat} coincides,
for small $\vec p_{i\perp}$, with that found using a non-conserved vector current model~\cite{a7}, which gives a
good description of the experimental data of the WA102 Collaboration~\cite{a2}. However, from the discussion
above, it is clear that these results simply follow from the general rules of Reggeon theory. They are not
connected with a particular vector current model of the Pomeron, but rather follow from the fact that the product
of the parity and signature of the Pomeron is $+1$. Moreover the Pomeron has positive signature and corresponds to
the analytic continuation from angular momenta $J^P = 2^+,4^+,\dots$ in the $t$-channel. The same results would be
obtained if a tensor current model were used.

On the other hand, the detailed structure of the amplitudes $f_m^k(p_{3\perp}^2, p_{4\perp}^2, \vec
p_{3\perp}\!\cdot\!\vec p_{4\perp})$ (with $m=0^+,0^-,1^+,2^+,\dots$) depends on dynamics and cannot be predicted
from the general principles of Regge theory. For example, if the heavy state $h$ is strongly coupled to gluons and
is produced perturbatively via the diagram shown in Fig.~\ref{fig:3}(a), then in the forward direction
($p_{3\perp}, p_{4\perp}\ll Q_\perp$) the vertex factor $f_{0^+}$ does not depend on $p_3$ or $p_4$. Moreover, as
was shown in \cite{Liverpool,KMRmm,INC}, there exists a $J_z=0$, parity-even, selection rule, for production by
gluon--gluon fusion where each of these active gluons comes from colour-singlet digluon $t$-channel exchange, see
Fig.~\ref{fig:3}. As a consequence the production of the negative-parity $h(0^-)$ state is strongly suppressed in
comparison with the production of the $h(0^+)$ state. Similarly it follows that the central diffractive exclusive
production of $2^{++}$ states is also suppressed in some topical cases; for example, $2^{++}$ states formed from
heavy quark pairs (in the non-relativistic model)~\cite{KMRmm,TEN} or $2^{++}$ `gravitons' in models with extra
dimensions in which their coupling to gluons has a point-like nature (with no derivatives) so they are not
produced via the $J_z=0$ two-gluon state~\cite{INC}. Also these processes can provide a unique opportunity to
determine the quantum numbers of pair-produced new strongly-interacting objects~\cite{INC}. For example,
comparatively light gluinos and squarks can be distinguished by the respective $\beta^3$ and $\beta$ threshold
behaviour, where $\beta$ is the particle velocity.

\section{Example: dynamics of $h(0^\pm)$ Higgs production} \label{sec:3}

So far we have discussed the structure of the production amplitudes for
\be pp \to p + h + p, \label{eq:rat1} \ee
where $h$ has a given $J^P$, which follow from general principles. To go further we need to consider the dynamics
of the process. We study $h(0^\pm)$ Higgs production as a topical example. The general rules imply that the
central vertices behave as
\be\begin{array}{l l l} g_{ik}^{h(0^+)} & \sim & {\rm constant} \nonumber\\
g_{ik}^{h(0^-)} & \sim & (\vec p_{3\perp} \times \vec p_{4\perp})\cdot \vec n_0 \ \sim
|t_1|^{\frac{1}{2}}|t_2|^{\frac{1}{2}}\sin\phi \label{eq:rat2} \end{array} \ee
at small $t$.

\subsection{Amplitudes for $h(0^{\pm})$ production}

\begin{figure}
\begin{center}
\includegraphics[height=5cm]{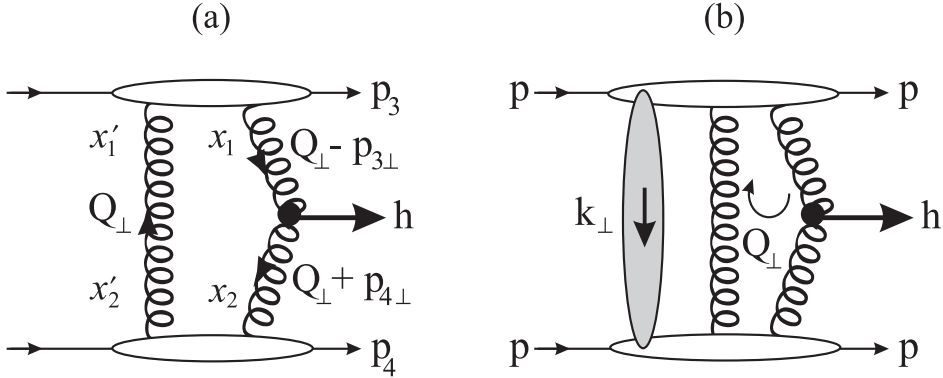}
\caption{(a)~The QCD diagram for double-diffractive exclusive production of a Higgs boson $h$, $pp\to p + h + p$,
where the gluons of the hard subprocess $gg\to h$ are colour screened by the second $t$-channel gluon.~~(b)~The
rescattering or absorptive corrections to $pp\to p + h + p$, where the shaded region represents the multi-Pomeron
exchanges of Fig.~\ref{fig:2}.\label{fig:3}}
\end{center}
\end{figure}

To see how the dynamics modify this behaviour we have first to describe how the cross sections for the exclusive
production of $h(0^\pm)$ Higgs bosons are calculated. We use the formalism of Ref.~\cite{KMR,KMRmm,INC}. The
amplitudes are described by the diagram shown in Fig.~\ref{fig:3}(a), where the hard subprocesses $gg\to h(0^\pm)$
are initiated by gluon--gluon fusion and where the second $t$-channel gluon is needed to screen the colour flow
across the rapidity gap intervals.
The Born amplitudes are of the form
\be T_h = N\int\frac{d^2Q_\perp\ V_h}{Q^2_\perp (\vec Q_\perp - \vec p_{3\perp})^2(\vec Q_\perp + \vec
p_{4\perp})^2}\: f_g(x_1, x_1', Q_3^2, \mu^2; t_1)f_g(x_2,x_2',Q_4^2,\mu^2; t_2) \label{eq:rat3} \ee
The overall normalization constant $N$ can be written in terms of the $h\to gg$ decay width~\cite{INC}, and the
$gg\to h$ vertex factors are
\be\begin{array}{l l l} V_{h(0^+)} & = & (\vec Q_\perp - \vec p_{3\perp}) \cdot (\vec Q_\perp + \vec p_{4\perp})\nonumber\\
V_{h(0^-)} & = & \left( (\vec Q_\perp - \vec p_{3\perp}) \times (\vec Q_\perp + \vec p_{4\perp})\right)\cdot \vec
n_0\,. \label{eq:rat4}
\end{array} \ee
The $f_g$'s are the skewed unintegrated gluon densities of the proton at the hard scale $\mu$, taken to be
$m_h/2$, with
\be\begin{array}{l l l} Q_3 & = & \min\left\{Q_\perp,|(\vec Q_\perp - \vec p_{3\perp})|\right\},  \nonumber\\
Q_4 & = & \min\left\{Q_\perp,|(\vec Q_\perp + \vec p_{4\perp})|\right\}. \label{eq:rat5}
\end{array} \ee
Below, we assume factorization of the unintegrated distributions,
\be f_g(x,x',Q^2,\mu^2;t) = f_g(x,x',Q^2,\mu^2)F_N(t), \label{eq:rat5a} \ee
where we parameterize the form factor of the proton vertex by the form $F_N(t) = \exp(bt)$ with $b=2~\GeV^{-2}$.
To single log accuracy, we have~\cite{MR01}
\be f_g(x,x',Q_i^2,\mu^2) = R_g\frac{\partial}{\partial \ln Q_i^2} \left(\sqrt{T(Q_i,\mu)}\: xg(x,Q_i^2)\right),
\label{eq:rat6} \ee
where $T$ is the usual Sudakov form factor which ensures that the gluon remains untouched in the evolution up to
the hard scale $\mu$, so that the rapidity gaps survive. The square root arises because the bremsstrahlung
survival probability $T$ is only relevant to the hard gluon. $R_g$ is the ratio of the skewed $x'\ll x$ integrated
distribution to the conventional diagonal density $g(x,Q^2)$. For $x\ll 1$ it is completely
determined~\cite{SGMR}. The apparent infrared divergence of~\eq{rat3} is nullified\footnote{In addition, at LHC
energies, the effective anomalous dimension of the gluon gives an extra suppression of the contribution from the
low $Q_\perp$ domain~\cite{KMRH}.} for $h(0^+)$ production by the Sudakov factors embodied in the gluon densities
$f_g$. However the amplitude for $h(0^-)$ production is much more sensitive to the infrared contribution. Indeed
let us consider the case of small $p_{i\perp}$ of the outgoing protons. Then, from~\eq{rat4}, we see that
$V_{h(0^+)} \sim Q_\perp^2$, whereas $V_{h(0^-)} \sim p_{3\perp}p_{4\perp}$ (since the linear contribution in
$Q_\perp$ vanishes after the angular integration). Thus the $d^2Q_\perp/Q_\perp^4$ integration for $h(0^+)$ is
replaced by $p_{3\perp}p_{4\perp} d^2Q_\perp/Q_\perp^6$ for $h(0^-)$, and now the Sudakov suppression is not
enough to prevent a significant contribution from the $Q_\perp^2\lesim1~\GeV^2$ domain.

\subsection{Uncertainties}

To estimate the uncertainty in the predictions for the $h^\pm(0)$ exclusive diffractive cross sections we first
quantify the above uncertainty arising from the infrared region, where the gluon distribution is not well known.
Fig.~4 shows the $\phi$ dependence of $h(0^-)$ and $h(0^+)$ production at the LHC, for $m_h=120$~GeV and
$\mu=m_h/2$, using different treatments of the infrared region.
\begin{figure}[!htb]\begin{center}
\epsfig{figure=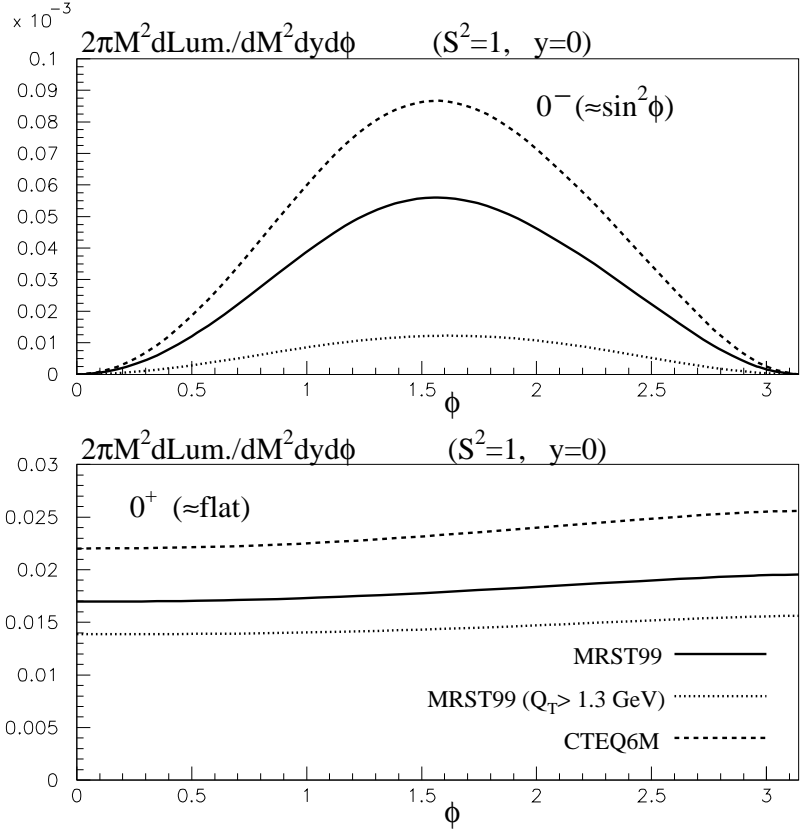,width=12cm}
 \caption{The $\phi$ dependences of diffractive exclusive $h(0^+)$ and $A(0^-)\equiv h(0^-)$ production, $pp\to p + h + p$, at the LHC, with $m_h = 120~\GeV$.
 $\phi$ is the angle in the transverse plane between the outgoing protons. The curves are for central rapidity and do
 not include absorptive corrections. They correspond to the effective luminosities for the $gg\to h(0^\pm)$
 subprocesses integrated over the outgoing proton momenta $\vec p_{i\perp}$, assuming an $\exp(-b p_{i\perp}^2)$
 behaviour of the unintegrated gluon densities $f_g$, with slope $b = 2~\GeV^{-2}$.} \label{fig:4} \end{center}
\end{figure}
The continuous and dashed curves are calculated using MRST99~\cite{MRST99} and CTEQ6M~\cite{CTEQ} partons
respectively with the very low $Q$ gluon frozen at its value at $Q_{3,4}=1.3$~GeV. Then we integrate down in
$Q_\perp$ until $Q_{3,4}$ are close to $\Lambda_{\rm QCD}$, where the contribution vanishes due to the presence of
the $T$-factor. This will slightly overestimate the cross sections as the gluon density decreases with decreasing
$Q^2$ for $x\sim0.01$. A lower extreme is to remove the contribution below $Q_{3,4}=1.3$~GeV entirely. The result
for MRST99 partons is shown by the dotted curve in Fig.~4. Even with this extreme choice, the $0^+$ cross section
is not changed greatly; it is depleted by about 20\%. On the other hand, as anticipated, we see for $0^-$
production, the infrared region is much more important and the cut reduces the cross section by a factor of 5.

Another uncertainty is the choice of factorization scale $\mu$. Note that in comparison with previous
calculations~\cite{INC}, which were done in the limit of proton transverse momenta, $p_{3,4\perp}\ll Q_\perp$, now
we include the explicit $p_\perp$-dependence in the $Q_\perp$-loop integral of (\ref{eq:rat3}). Moreover, we resum
the `soft' gluon logarithms, $\ln\,1/(1-z)$, in the $T$-factor.\footnote{To account for the interference and the
precise form of the amplitude for soft gluon ($q_\perp \ll m_h$) large-angle emission, we explicitly calculate the
one-loop virtual correction to the $gg\to h$ vertex, integrating over the whole angular range for $q_\perp \ll
m_h$. We adjust the upper limit of the $z$-integral so that $z<0.62m_h/(0.62m_h+q_\perp)$ in the expression for
the $T$-factor (see eq.~(10) of \cite{INC} with $k_t=q_\perp$), in order to reproduce the complete one-loop
result.}
So now the $T$-factor includes both the single soft logarithms and the single collinear logarithms. The only
uncertainty is the non-logarithmic NLO contribution. This may be modelled by changing the factorization scale,
$\mu$, which fixes the maximal $q_\perp$ of the gluon in the NLO loop correction. As the default we have used
$\mu=m_h/2$; that is the largest $q_\perp$ allowed in the process with total energy $m_h$. Choosing a lower scale
$\mu=m_h/4$ would enlarge the cross sections by about 30\%.

Next there is some uncertainty in the gluon distribution itself. To illustrate this, we compare predictions
obtained using CTEQ6M~\cite{CTEQ}, MRST99~\cite{MRST99} and MRST02~\cite{MRST02} partons. For $0^+$ production at
the LHC, with $m_h=120$~GeV and $\mu=m_h/2$, we find that the effective gluon--gluon luminosity, before screening,
is
\begin{equation}
\left.\frac{d{\cal L}}{dyd\ln M^2}\right|_{y=0} = (2.2,\ 1.7,\ 1.45)\times10^{-2}
\end{equation}
respectively. This spread of values arises because the CTEQ gluon is 7\% higher, and the MRST02 gluon 4\% lower,
than the default MRST99 gluon, in the relevant kinematic region.  The sensitivity to the gluon arises because the
central exclusive diffractive cross section is proportional to the 4th power of the gluon. For $0^-$ production,
the corresponding numbers are $(4.2,\ 2.7,\ 1.7)\times10^{-5}$. Up to now, we have discussed the effective
gluon--gluon luminosity. However, NNLO corrections may occur in the $gg\to h$ fusion vertex. These give an extra
uncertainty of $\pm20\%$. Note that we have already accounted for the NLO corrections for this vertex~\cite{INC}.

Finally, we need to consider the uncertainty in the evaluation of the soft rescattering correction factor $S^2$,
which is the probability that the rapidity gaps survive the soft $pp$ interaction. The computation of $S^2$ is
discussed in some detail in Section~4. Here it suffices to say, from the analysis~\cite{KMRsoft} of all soft $pp$
data, that a conservative error on the values of $S^2$ is $\pm50\%$.

Combining together all these sources of error we find that the prediction for the $0^+$ cross section is uncertain
to a factor of almost 2.5, that is up to almost 2.5, and down almost to $1/2.5$, times the default
value.\footnote{For example, we predict the cross section for the exclusive diffractive production of a Standard
Model Higgs at the LHC, with $m_h = 120$~GeV, to be in the range 0.9--5.5~fb.}
On the other hand, $0^-$ production is uncertain by this factor just from the first (infrared) source of error,
with the remaining errors contributing almost another factor of 2.5. Although the rate of $h(0^-)$ production is
very sensitive to the infrared contribution, and could indicate the presence of a significant non-perturbative
contribution, we find that the form of the $\phi$ dependence is not. We discuss this point in Section~\ref{sec:5}.

Note that the non-local structure of the amplitude leads to an extra angular dependence coming from the
correlations between $\vec Q_\perp$ and the $\vec p_{i\perp}$ in the integral in~\eq{rat3}. In fact, expanding the
gluon propagators gives corrections of the type
\be \vec Q_\perp \cdot \vec p_{3\perp} / Q_\perp^2,\qquad \vec Q_\perp \cdot \vec p_{4\perp} / Q_\perp^2,
\label{eq:rat7} \ee
which lead to an additional contribution of the form $-\vec p_{3\perp}\cdot \vec p_{4\perp}/Q_\perp^2$. This
reflects the dependence of the vertex factors $f_{0^\pm}$ on $\vec p_{3\perp}\cdot \vec p_{4\perp}$,
see~\eq{nightingale} and \eq{ocelot}. However, it is evident from Fig.~\ref{fig:4} that this contribution does not
give a large effect. What is more important is the suppression of $h(0^-)$ production in comparison to that for
$h(0^+)$. The $h(0^-)$ amplitude is proportional to $\vec p_{3\perp}\times \vec p_{4\perp}$, where the dimensions
must be compensated by some scale. In perturbative QCD this is the  scale $Q_\perp^2$ arising from the gluon loop
in Fig.~\ref{fig:3}(a). Therefore the $h(0^-)$ cross section is reduced by a factor $\langle\, p_{3\perp}^2
p_{4\perp}^2 / 2Q_\perp^4\,\rangle$, that is by a factor of the order of 500 for typical $p_{i\perp}^2 \sim 1/2b
\sim 0.25~\GeV^2$ if $Q_\perp^2 \sim 4~\GeV^2$.

\section{Absorptive corrections}\label{sec:4}

In this section we consider how exclusive double-diffractive production is influenced by the absorptive
(shadowing) effects, which arise from the multi-Pomeron diagrams of Fig.~\ref{fig:2}. To determine the suppression
due to these absorptive corrections, it is convenient to work in impact parameter, $b$, space.

\subsection{Absorptive effects for $h(0^+)$ production}

The amplitude for the central production of an $h(0^+)$ state, via the double-Pomeron-exchange process $pp\to p +
h + p$, has the form
\be T^h(s,\vec b_1,\vec b_2,\vec b) = \exp\left({-\frac{1}{2}\Omega_P(s,b^2)}\right) T_{PP}^h(s_1,s_2,\vec
b_1,\vec b_2), \label{eq:snake} \ee
where $\Omega_P$ is the contribution of Pomeron exchange to elastic $pp$ scattering in impact parameter space
\be \Omega_P(s,b^2) =  \frac{\sigma_{pp}^P}{4\pi B} \exp (-b^2/4B), \label{eq:tiger} \ee
where $\sigma_{pp}^P$ is the Pomeron contribution to the total cross section of the $pp$ interaction, and
\be B = {\textstyle \frac{1}{2}}B_0 + \alpha_P'\ln(s/s_0) \label{eq:unicorn} \ee
is the slope of the elastic $pp$ amplitude. The amplitude $T_{PP}^h$ is the Fourier transform, to impact-parameter
space, of the amplitude $T_{12}^{3h4}(s_1,s_2,\vec p_{3\perp},\vec p_{4\perp})$ of~\eq{adder} with $i=k=P$. That
is
\be  T_{PP}^h(s_1,s_2,\vec b_1,\vec b_2) = \left(\frac{1}{2\pi}\right)^2\int d^2p_{3\perp}d^2p_{4\perp}\: e^{i\vec
p_{3\perp}\cdot\vec b_1} e^{-i\vec p_{4\perp}\cdot \vec b_2}\: T_{PP}^h(s_1,s_2,\vec p_{3\perp},\vec p_{4\perp}),
\label{eq:vole} \ee
where $\vec b = \vec b_1 + \vec b_2$ is the Fourier conjugate to $\vec q = \vec p_{3\perp} - \vec p_{4\perp}$. For
simplicity, we give the formula for a single-channel eikonal, where only intermediate proton states are
considered, between the Pomeron exchanges in Fig.~\ref{fig:2}. The extension to the multichannel case is
straightforward. In the calculations presented here we used the two-channel eikonal model of Ref.~\cite{KMRsoft}.

Note that if $\alpha_P(0)-1 \equiv \Delta>0$, then $\Omega_P(s,b^2)$ increases with energy and leads to a
substantial suppression of cross section at very high energies. Calculations, using the model of
Ref.~\cite{KMRsoft}, show that at Tevatron energies the Born cross section, corresponding to the diagram of
Fig.~\ref{fig:1}(b), is suppressed by the multi-Pomeron exchanges of Fig.~\ref{fig:2} by a factor of roughly 0.05.
At the LHC the suppression factor\footnote{It is interesting to note that the introduction of the $\vec
p_{i\perp}$ and angular correlations in (\ref{eq:rat3}), (\ref{eq:rat4}) raise $\langle S^2\rangle$ from 0.023 to
0.026.}
$\langle S^2\rangle$ is 0.026.

Since the amount of suppression depends on the impact parameter $\vec b$, it leads to a characteristic dependence
of the factor $S^2$ on the angle $\phi$ between the outgoing protons~\cite{a17}. This is related to the fact that
$\vec b$ is the Fourier conjugate to the vector $\vec q = \vec p_{3\perp} - \vec p_{4\perp}$. If the outgoing
protons are tagged, then the characteristic peripheral form of the amplitude $T^h$ in $\vec b$-space can be
studied experimentally in double-Pomeron-exchange processes by measuring the dependence of the cross section on
$\vec q$.

We emphasize that the suppression $S^2$, due to absorptive or rescattering corrections, depends not only on the
particular process, but also on the kinematical cuts which select events in a given $p_{i\perp},\phi$ domain.
Therefore the suppression $S^2$ has to be calculated for each particular kinematical configuration.

\subsection{Comparison of exclusive diffractive $h(0^\pm)$ Higgs production} \label{sec:4a}

So far we have discussed absorptive corrections for $h(0^+)$ production. Here we compare these corrections with
those for $h(0^-)$ production. For $h(0^-)$ production we predict a different $\vec q$ behaviour. This originates
from~\eq{ocelot}; the Born double-Pomeron-exchange amplitude for process~\eq{adder} now contains the kinematical
factor $(\vec p_{3\perp}\times \vec p_{4\perp})\cdot \vec n_0$ and this, in turn, implies that the Fourier
transform contains the factor $(\vec b_1 \times \vec b_2)\cdot \vec n_0$. Thus the corresponding amplitude
$T_{PP}^{h(0^-)}(s_1,s_2,\vec b_1,\vec b_2)$ tends to zero as $\vec b_1$ or $\vec b_2\to0$. As a result, the
suppression arising from rescattering is less effective, and the factor $S^2$ is larger than for $h(0^+)$
production. Also the $\phi$ distribution is distorted due to absorption.

The effect of the absorptive corrections on the angular correlations $\phi$ between the outgoing protons was
discussed in detail in Ref.~\cite{a17} for $h(0^+)$ production\footnote{Note that there is a typographical error
in eq.~(25) of \cite{a17}, where the last factor should be simply $S^2$ rather than its second derivative. However
the results presented in \cite{a17} correspond to the correct definition of $F$.}. There it was shown that the
absorptive corrections are largest in the back-to-back configuration where $\vec p_{3\perp}$ is directed against
$\vec p_{4\perp}$, since in this case both $t_1 \simeq -(\vec k_\perp + \vec p_{3\perp})^2$ and $t_2\simeq -(\vec
k_\perp - \vec p_{4\perp})^2$ can be minimized simultaneously by the same momentum $\vec k_\perp$ transferred in
the elastic rescattering, see Fig.~\ref{fig:3}(b). Thus for $\phi = 180^\circ$ the momentum is transferred mainly
through the rescattering amplitude. The suppression factor $S^2$ was plotted versus $\phi$ for different choices
of $p_{3\perp}$ and $p_{4\perp}$ in Ref.~\cite{a17}. It was shown that the diffractive dip (which arises from the
maximum cancellation between the bare amplitude and rescattering contribution) moves to smaller $\phi$ as the
values of $p_{i\perp}$ are increased.

\begin{figure}[!htb]\begin{center}
\epsfig{figure=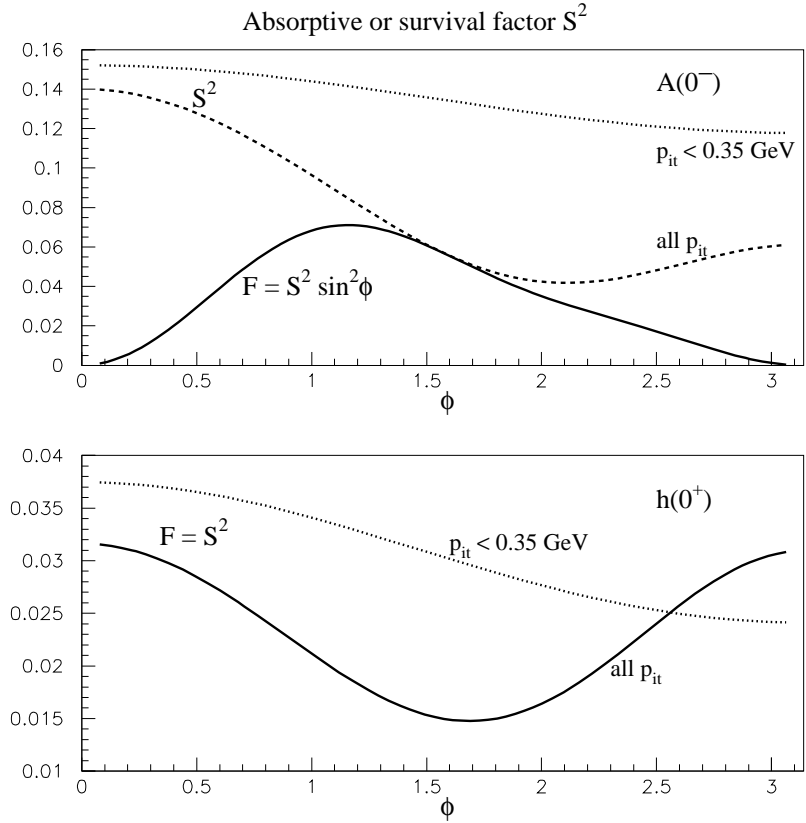,width=12cm}
 \caption{The suppression factors $S^2$ of $h(0^+)$ and $A(0^-)\equiv h(0^-)$ Higgs production via the process $pp\to p + h + p$ at the LHC, arising
 from rescattering effects. The outgoing protons are integrated over (i)~all $p_{i\perp}$ and (ii)~$p_{i\perp} <
 0.35~\GeV$ (dotted curves). For illustration, the continuous curve for $h(0^-)\equiv A(0^-)$ production includes the general
 $\sin\phi$ behaviour of the bare amplitude.} \label{fig:5} \end{center}
\end{figure}

Here we calculate $S^2$ as a function of $\phi$ for $h(0^-)$, as well as $h(0^+)$, exclusive diffractive
production. We integrate over the $p_{i\perp}$ of the outgoing protons assuming an $\exp(-b(p_{3\perp}^2 +
p_{4\perp}^2))$ behaviour of the Pomeron-proton vertices $g_{13}$ and $g_{24}$, with $b=2~\GeV^{-2}$. We use the
two-channel eikonal model of Ref.~\cite{KMRsoft}. For the central vertex we take $(\vec p_{3\perp} \times \vec
p_{4\perp})\cdot \vec n_0$ for $h(0^-)$ production and a constant for $h(0^+)$ production. The results for the
suppression factor $S^2$ are shown in Fig.~\ref{fig:5} for $h(0^\pm)$ production of mass 120~GeV at the LHC
energy, $\sqrt s = 14$~TeV.
As the azimuthal angle between $\vec p_{3\perp}$ and $\vec p_{4\perp}$ increases, the first diffractive dip,
followed by the second maximum, are evident in the $S^2$ curves obtained by integrating over all $p_{i\perp}$. The
dotted curves show the effect of restricting the outgoing protons to the domain $p_{i\perp}<0.35~\GeV$. As
expected, the diffractive dip is pushed to larger angles and is barely reached even for the back-to-back
configuration, $\phi=180^\circ$. As we see from Fig.~\ref{fig:5} that the survival factor $S^2$ is about 3--4
times larger for $h(0^-)$ as compared to $h(0^+)$ production. This is a reflection of the more peripheral nature
of $h(0^-)$ production. For the same reason the suppression $S^2$ obtained when integrating over the small
$p_{i\perp}$ domain, $p_{i\perp}<0.35~\GeV$, is less than when integrating over all $p_{i\perp}$, since it is more
weighted to the larger values of the impact parameter $b$.

\begin{figure}[!htb]\begin{center}
\epsfig{figure=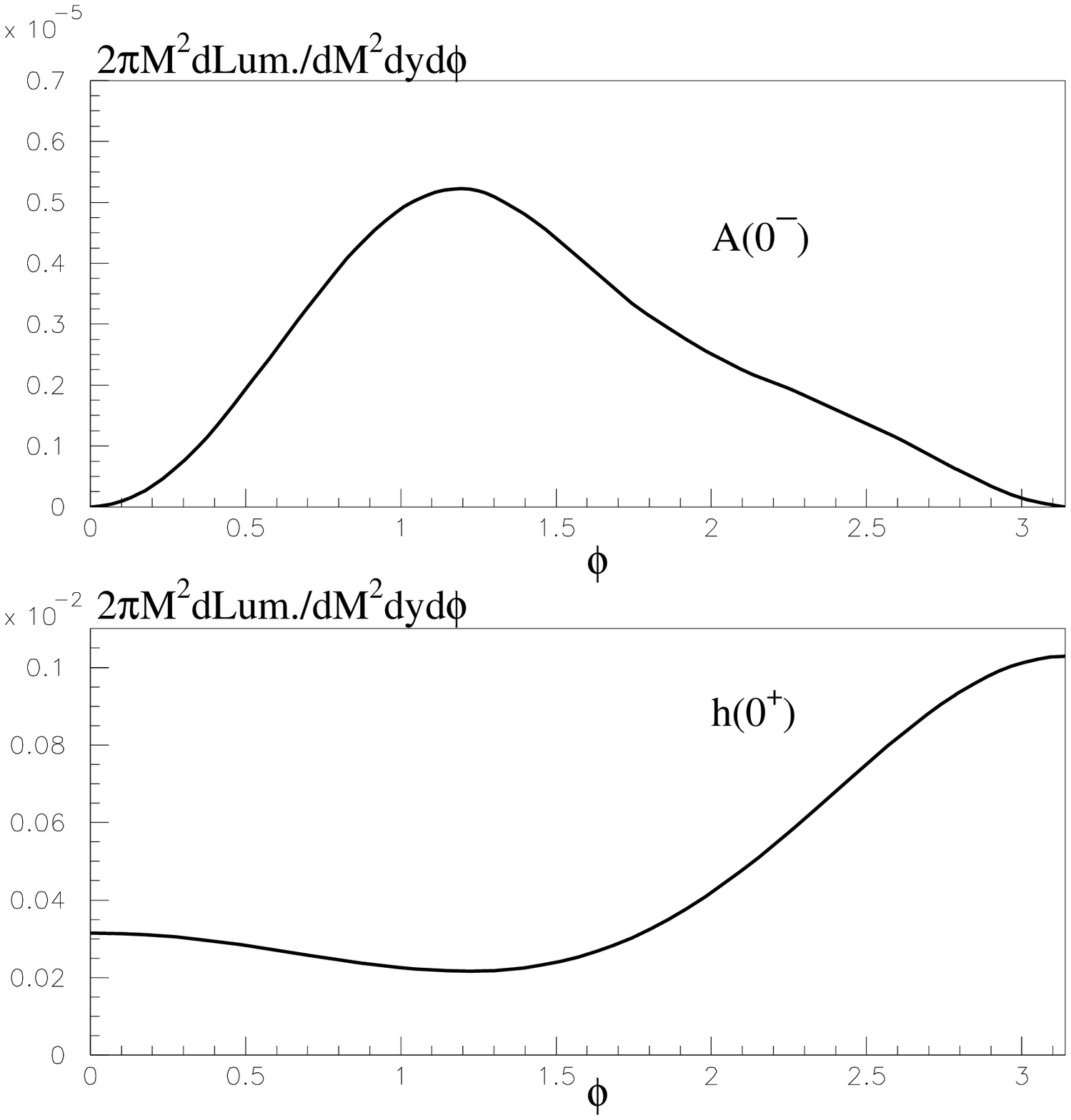,width=12cm}
 \caption{The effective luminosities of the $gg\to A(0^-)\equiv h(0^-)$ and $h(0^+)$ subprocesses, integrated over the outgoing proton
 transverse momenta $\vec p_{i\perp}$, in the exclusive diffractive processes $pp\to p + h + p$ at the LHC, with $\sqrt s = 14$~TeV, using the MRST99~\cite{MRST99} gluon.
 Absorptive (or rescattering) effects are included.  The curves are for $y=0$ and Higgs masses of 120~GeV.
 Comparison with the continuous curves in Fig.~\ref{fig:4} shows the
 suppression of the event rate, and the distortion of the $\phi$ behaviour due to absorptive effects.} \label{fig:6}\end{center}
\end{figure}
Finally in Fig.~\ref{fig:6} we show the predictions for the effective luminosity with the absorptive effects
included. The original $\sin^2\phi$ and constant behaviours of $h(0^-)$ and $h(0^+)$, respectively, are distorted
first by the $\vec p_{3\perp}\cdot \vec p_{4\perp}/Q^2$ type corrections from the integration over the gluon loop
in Fig.~\ref{fig:4}, and then by the absorptive effects given by the suppression factors $S^2$ shown in
Fig.~\ref{fig:5}.

\section{Consequences for signals in the Higgs sector}\label{sec:5}

We have studied the central {\em exclusive} diffractive production of bosons via the process $pp\to p + h + p$,
and emphasized that correlations between the outgoing proton momenta reflect the spin-parity of $h$. As a topical
example to illustrate these properties we compared $h(0^+)$ and $A(0^-)\equiv h(0^-)$ Higgs production. In
particular, Fig.~\ref{fig:6} shows that the dependence on the angle $\phi$ between the outgoing proton transverse
momenta, $\vec p_{3\perp}$ and $\vec p_{4\perp}$, is different for the natural ($0^+$) and unnatural ($0^-$)
parity states of $h$. The comparison with Fig.~\ref{fig:4} shows that absorptive effects have significantly
distorted the $\phi$ distributions and, in fact, have increased the difference between the $0^+$ and $0^-$
distributions. Thus this distribution provides a unique possibility to distinguish between $0^+$ and $0^-$ bosons,
which, in the case of {\em inclusive} production\footnote{A proposal, similar in spirit to our approach, can be
found in Ref.~\cite{OZ}. The idea is to determine the CP-parity of a Higgs boson by measuring the azimuthal
correlations of the tagging (quark) jets which accompany Higgs production via the vector-boson-fusion mechanism.
Even if we disregard the possible degradation of the characteristic features of the distribution caused by parton
showers and the inclusive environment of the jets, we note that the method is not applicable in some important
regions where the couplings of the Higgs to vector bosons are strongly suppressed. Another method to determine the
spin-parity of the Higgs, which similarly relies on the vector-boson coupling, was discussed in
Ref.~\cite{CMMZ}.}, is extremely difficult.

We have seen that the amplitude for the production of unnatural parity ($P=(-1)^{J+1}$) states contains a factor
$(\vec p_{3\perp}\times\vec p_{4\perp})\cdot\vec n_0$. Thus the cross section vanishes as $p_{3\perp}$ or
$p_{4\perp}\to0$ and vanishes as $\sin^2\phi$ as $\phi\to0$ or $\pi$. These properties may be used to suppress the
cross section for natural parity ($P=(-1)^J$) production in comparison to that of unnatural parity states. In
particular, selecting events with $p_{3\perp},p_{4\perp}>0.4~\GeV$ and $20^\circ < \phi < 120^\circ$ suppresses
the $0^+$ yield by about a factor of 10, while only decreasing the $0^-$ cross section by a factor of 2.3. The
relative $0^-$ enhancement may be important as the cross section for the central exclusive production of an
$A(0^-)$ boson is quite small, and moreover, in many supersymmetric scenarios, the $h(0^+)$ (and/or $H(0^+)$) and
$A(0^-)$ bosons are close in mass.

\begin{figure}[!htb]\begin{center}
\epsfig{figure=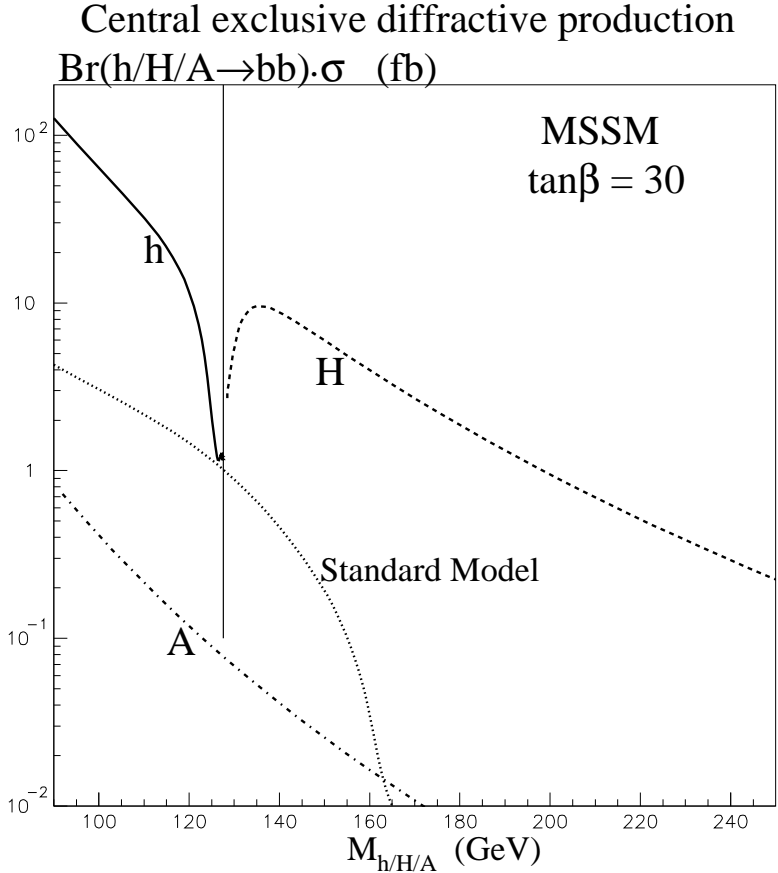,width=12cm}
 \caption{The cross sections times the $\bb$ branching ratio predicted, for the central exclusive production of
 $h(0^+)$, $H(0^+)$ and $A(0^-)$ MSSM Higgs bosons (for $\tan\beta = 30$), at the LHC, compared with the SM
 result. We use the MRST99~\cite{MRST99} gluon distribution. The vertical line separates the mass regime of light $h(0^+)$ and heavy $H(0^+)$ bosons.} \label{fig:7}\end{center}
\end{figure}

As a specific example we consider Higgs production in the minimal supersymmetric model (MSSM) with large
$\tan\beta$ and $m_A \sim 110$--130~GeV.  In this domain the branching ratios of Higgs-like bosons to vector
bosons and photon pairs\footnote{The branching ratios of $h,H,A\to\gamma\gamma$ are less than, or of the order of,
$10^{-5}$--$10^{-4}$, which is much smaller than in the SM.}, and the couplings to top quarks, are much
suppressed~\cite{EB}, and it becomes problematic to perform a complete coverage of the Higgs boson sector using
the conventional inclusive processes. In particular the problem of resolving the signals for different states
becomes quite challenging\footnote{The separation of $h$ and $H$ bosons may be especially difficult for {\em
inclusive} signals, where the mass resolution is usually $\Delta m\gtrsim 10~\GeV$, except in the $\gamma\gamma$
and probably $\mu\mu$ modes. However, with forward proton taggers, the {\em exclusive} signal has the added bonus
that a mass resolution of $\Delta m \sim 1~\GeV$ may be obtained~\cite{DKMOR}.}.  On the contrary the cross
section, $\sigma_{\rm CEP}$, for central exclusive diffractive production in the MSSM is enhanced in comparison
with that of the SM. The MSSM (for $\tan\beta = 30$) and SM cross sections, $\sigma_{\rm CEP}$, at the LHC energy,
are shown in Fig.~\ref{fig:7}.
They have been evaluated using the effective $gg$ luminosities obtained in Section~\ref{sec:3} with the absorptive
corrections calculated in Section~\ref{sec:4a}, see Fig.~\ref{fig:6}. The normalization factor in~\eq{rat3} is
\be N^2 = \frac{2\pi\Gamma(h\to gg)}{(N_c^2 - 1)^2 m_H^3}\,K, \label{eq:wombat} \ee
where the NLO $K$ factor $\simeq 1.5$~\cite{INC} and the number of colours $N_c = 3$. The widths and properties of
the Higgs scalar ($h,H$) and pseudoscalar ($A$) bosons are calculated using the HDECAY code, version
3.0~\cite{HDEC}, with all other parameters taken from Table~2 of \cite{HDEC}; also we take ${\rm IMODEL} = 4$,
which means the radiative corrections are included according to Ref.~\cite{HHW}.

From Fig.~\ref{fig:7} we see that the $0^+$ Higgs bosons should be accessible at the LHC, via the central
exclusive signal, over a wide mass range up to about 250~GeV in this scenario. The enhancement of the MSSM signals
is clearly apparent, except near $m_h\simeq 127~\GeV$. For example, for the production of a Higgs boson of mass
115~GeV for $\tan\beta = 30$ (or 50) we have
\be {\rm Br}(h\to\bb)\:\sigma_{\rm CEP}^{h(0^+)}\ \sim\ 20~(70)~{\rm fb}, \label{eq:xrayfish} \ee
about 10~(40) times larger than $\sigma_{\rm CEP}$ in the SM.

For the same parameters, for $A(0^-)$ production we obtain
\be {\rm Br}(A\to\bb)\:\sigma_{\rm CEP}^{A(0^-)}\ \sim\ 0.2~(0.5)~{\rm fb}. \label{eq:yak} \ee
However we have emphasized the infrared sensitivity of the {\em rate} of $A(0^-)$ production. It is possible that
a non-perturbative contribution, coming from low values of $Q_\perp$ in~\eq{rat3}, may enhance the cross section
by a factor of 3 or more. Nevertheless it would be extremely hard to observe the $A(0^-)$ boson under the $h(0^+)$
signal, when the masses are close. A typical mass difference is $m_A-m_h \simeq 2.9$~(1.4)~GeV for $\tan\beta =
30$~(50), if $m_A = 115~\GeV$. The situation for the observation of the $A(0^-)$ boson is even worse due to the
comparatively large expected widths of the Higgs bosons. For instance, if $m_A\sim 90\!\!-\!\!130~\GeV$ and
$\tan\beta = 30$, then the widths of the Higgs bosons become of order 2~GeV~\cite{CH}. On the other hand, proton
taggers, with a very accurate missing mass resolution of $\Delta M \simeq 1~\GeV$, offer the attractive
possibility, not only to separate the $h$ and $H$ bosons, but also to provide a direct measurement of the widths
of the $h$ (for $m_h\lesim 120~\GeV$) and the $H$ bosons (if $m_H\gtrsim 130~\GeV$). Also we note that by
comparing the cross sections of~\eq{xrayfish} and \eq{yak}, we see that if a new heavy object were observed in
inelastic events, but not in exclusive central production, it would indicate that it had unnatural parity.

Although the rate of $A(0^-)$ Higgs production is sensitive to contributions from the infrared region, we do not
expect a significant change in the qualitative behaviour of the $0^-$ production amplitude. The reasons are as
follows. First, the vanishing of the amplitude as $p_{i\perp}\to 0$ and/or $\phi\to 0,\pi$ follows from the
general form \eq{ocelot} of the vertex in Regge theory. Second, as a rule the extra $\phi$ dependence caused by
the $\vec p_{3\perp}\cdot\vec p_{4\perp}$ term is weak, see Fig.~4 for example. Third, in the very extreme case
where we use GRV94 partons~\cite{GRV94} (which enable us to take a very low infrared cut-off $Q_0^2 = 0.4~\GeV^2$,
but which are known to overestimate significantly the low $x$ gluon), the $0^-$ Higgs cross section is enhanced,
relative to that obtained using MRST99 partons with $Q_0=1.3~\GeV$, by about a factor of 4, but the $\phi$ and
$p_{i\perp}$ dependences are essentially unaltered.

Returning to $h(0^+)$ production, we see, for the example of~\eq{xrayfish}, that already for an LHC luminosity
${\cal L} = 30~{\rm fb}^{-1}$, about 600~(2000) bosons are produced. If the experimental cuts and efficiencies
quoted in Ref.~\cite{DKMOR} are imposed, then the signal is depleted by about a factor of 6. This leaves about
100~(400) observable events, with an unaltered background of about 3~events~\cite{DKMOR} in a $\Delta M = 1~\GeV$
missing mass bin; which gives an incredible significance for a Higgs signal!

%%%%%%%%%%%%%%%%%%%%%%%%%%%%%%%%%%%%%%%%%%%%%%%%%%%%%%%%%%%%%%%%%%%%%%%%%%%

%%%%%%%%%%%%%%%%%%%%%%%%%%%%%%%%%%%%%%%%%%%%%%%%%%%%%%%%%%%%%%%%%%%%%%%%%%%

%\newpage
\section*{Acknowledgements}

We thank Abdelhak Djouadi, Howie Haber, Leif Lonnblad, Risto Orava, Albert de Roeck and Georg Weiglein for
valuable discussions. ABK and MGR would like to thank the IPPP at the University of Durham for hospitality. This
work was supported by the UK Particle Physics and Astronomy Research Council, by grants INTAS 00-00366, RFBR
01-02-17383 and 01-02-17095, and by the Federal Program of the Russian Ministry of Industry, Science and
Technology 40.052.1.1.1112 and SS-1124.2003.2.

%\newpage

%%%%%%%%%%%%%%%%%%%%%%%%%%%%%%%%%%%%%%%%%%%%%%%%%%%%%%%%%%%%%%%

%%%%%%%%%%%%%%%%%%%%%%%%%%%%%%%%%%%%%%%%%%%%%%%%%%%%%%%%%%%%%%%

\end{document}